\documentclass{article}    

\usepackage{graphicx}

\title{\textbf{Rule 4 Statistics}}  
\author{Richard Mould\footnote{Department of Physics and Astronomy, State University of New York, Stony Brook,
\mbox{New York} 11794-3800; http://nuclear.physics.sunysb.edu/ \~{}mould}}  
\date{}    
   
\begin{document}             

\maketitle              

\begin{abstract}

When a conscious observer is part of a quantum mechanical system, rule (4) cuts off solutions to the Schršdinger
equation.  It is important to show that this interruption of the Hamiltonian dynamics does not effect the statistical
predictions of the theory.  The initial case considered is that of a two atom radioactive source.  It is found that when
the predictions of standard (Born rule) quantum theory are verified by using a particular experimental procedure, the
result is the same as that predicted by quantum theory qualified by rule (4).  This example is generalized, and the result
is found to be the same.

\end{abstract}

\section*{Introduction}

 		Standard quantum theory (i.e., the equations of motion plus Born's interpretation) is limited to ensembles of quantum
mechanical events.  It cannot be generally applied to individual members of an ensemble.  The theory is furthermore
limited in that it cannot be applied to systems that include a conscious observer.  However, it is found in previous
papers \cite{RM1, RM2} that when the Born rule is discarded and other rules are put in its place, quantum mechanical
systems \emph{can} include conscious observers, and the theory \emph{can} be applied to individual cases.  

	Four new rules are required to accomplish this broadening of quantum theory.  The fourth of these rules is a selection
rule on brain states that disrupts the flow of probability current to second order transitions.  It would seem that this
interruption of the smooth application of Hamiltonian dynamics would have an adverse experimental effect.  It would seem
that rule (4) would skew the statistics of a quantum mechanical system in which a conscious observer is present.  However,
it is shown in this paper that that is not the case.  It is found that the theoretical amendment imposed by rule (4)
actually mimics a valid experimental procedure that can be used to verify the statistics.  The predictions of the theory
(including rule 4) are therefore found to be independent of the presence of a conscious observer.  

Rule (4) does not contradict the predictions of standard theory, and it allows those predictions to be correct when an
observer is continuously present \emph{inside} the system.

\section*{A Two-Atom Radioactive Source}

	Consider a two-atom radioactive source where each atom is initially in a state $a(t)$, and decays to a state $a_0(t)$. 
The final state $a_0$ is assumed to include the emitted (now free) particle.  The initial state of the system is then
written
\begin{displaymath}
\phi = (a + a_0)(a + a_0)d
\end{displaymath}
where $d$ is a detector that is not yet interacting with the source.  In any product such as $aa$, it is assumed that the
first $a$ is a state of the first atom, and the second $a$ is a state of the second atom.  The square modulus is given by
\begin{displaymath}
\phi^*\phi = (a^*a + a_0^*a_0)(a^*a + a_0^*a_0)d^*d
\end{displaymath}
which will be written
\begin{displaymath}
\Phi = (A + A_0)(A + A_0)D
\end{displaymath}
where $\Phi(t) = \phi^*\phi, A(t) = a^*a, A_0(t) = a_0^*a_0$, and $D = d^*d$.  Again, the first $A$ in a product like $AA$
refers to the square modulus of the first atom, and the second refers to the second atom.  

Expanding gives
\begin{equation}
\Phi = [AA + (AA_0 + A_0A) + A_0A_0]D
\end{equation}
so when the detector becomes entangled with the emitted particle in each case, we have
\begin{equation}
\Phi = AAD_0 + (AA_0 + A_0A)D_1 + A_0A_0D_2
\end{equation}
where $D_0$ is the detector with no counts, $D_1$ is the detector with 1 count, and $D_2$ is the detector with 2 counts. 
In the entangled form of eq.\ 2, the emitted particle is no longer assumed to be a free particle associated with $A_0$. 
It is captured by the detector.  

Setting $A = e^{-kt}$, and $A_0 = (1 - e^{-kt})$, it is clear that each atom remains normalized in time.  Applying these
values to eq. 1 gives
\begin{equation}
\Phi(t) = [e^{-2kt} +2e^{-kt}(1 - e^{kt}) + (1 - e^{kt})^2 ]
\end{equation}
inasmuch as $D$ has a square modulus equal to 1.0.  The first component in \mbox{eq.\ 3} is the rate of decay of the
radioactive source.  The second component $2e^{-kt}(1 - e^{-kt})$ is the rate at which a single particle is captured, and
the third component $(1 - e^{-kt})^2$ is the rate at which two particles are captured.  For very small times given by
$\epsilon = kt$, eq. 3 becomes  
\begin{displaymath}
\Phi(t=0) = \{1 - 2\epsilon\} + \{ 2\epsilon \} + \{ 0 \} = 1
\end{displaymath}
The first component initially looses square modulus in the amount $2\epsilon$, and this goes entirely into the second
component.  The third component is not initially affected.  This is because the Hamiltonian of the system does not provide
a direct connection between the first and the third components.

The probability current flowing into each of these components is given by
\begin{displaymath}
J(t) = d\Phi/dt =-2ke^{-2kt} + \{ -2ke^{-kt} + 4ke^{-2kt}\} + \{ 2ke^{-kt} - 2ke^{-2kt}\} = 0
\end{displaymath}
The initial current flow at $t = 0$ goes into the second component only
\begin{displaymath}
J(t= 0)/2k = -1 + 1 + 0 = 0
\end{displaymath}
Again, current flows only from the first to the second component.  It will not go to the third component until the second
has acquired some amplitude.

\section*{Empirical Verification}

The statistical outcome in eq.\ 3 must be empirically verified.  Imagine that the observer keeps an eye on the detector
from the time $t_0$ that he starts the clock.  He marks the time $t_1$ when the first capture occurs, at which time he
zeros the clock so it can record the time $t_2$ between the first and the second count.  When the second count occurs, he
again zeros the clock so it can record the time $t_3$ between the second and the third count.  This process is continued
until a time $t_n$ has been reached such that $\Sigma_nt_n \le t < \Sigma_nt_{n+1}$.  The detector will then
record $n$ counts at the time $t$.  Repeating this process many times establishes the distribution of counts that can be
found at time $t$.  This should be the same as the distribution given by eq.\ 3 at time $t$.  Stopping and starting the
clock in this way may seem to be an unnecessary complication.  However, it is procedurally correct, and it parallels the
action of rule (4) in refs.\ 1, 2. 

\section*{Continuous Observation - Rule (4)}

Let the above experiment by continuously observed by an experimenter.    Before the first particle capture at time
$t_1$ (i.e., before the first stochastic hit), eq.\ 2 is amended to read
\begin{equation}
\Phi(t_1 >t \ge 0) = AAD_0\underline{B}_0 + (AA_0 + A_0A)D_1B_1 + A_0A_0D_2B_2
\end{equation}
where the second and third components involving the experimenter's brain states $B_1$ and $B_2$ are equal to zero at $t =
0$.  The underlined state $\underline{B}_0$ is a conscious state, and the non-underlined states $B_1$ and $B_2$ are ready
brain states.  

	Rule (4) explicitly forbids current flow from one ready brain state to another, so there can be no current flow from the
second to the third component in \mbox{eq.\ 4}; and since there is no current flow from the first to the third component,
it follows that the third component in eq.\ 4 is not in the picture.  Therefore, eq.\ 4 takes the simpler form 
\begin{equation}
\Phi(t_1 >t \ge 0) = AAD_0\underline{B}_0 + (AA_0 + A_0A)D_1B_1
\end{equation}
Consequently, current leaving the first component can only go into the second component, giving
\begin{displaymath}
J = -2ke + 2ke = 0
\end{displaymath}
This results in a probability of 1.0 that $B_1$ in eq.\ 5 will be stochastically chosen, insuring that the experimenter 
can measure the time $t_1$.

At the moment of a stochastic hit on the ready brain state $B_1$, rule (3) requires a state reduction in which the first
component goes to zero, and the ready brain state $B_1$ becomes a conscious brain state $\underline{B}_1$.  Also at that
moment, there is an `effective' renormalization that comes about because rule (1) requires subsequent probability current
flow to be divided by the new square modulus.  Therefore, to the experimenter, a new (renormalized) cycle of observation
begins at $t_1$. 

Current will then flow exclusively into the ready brain state $B_2$, inasmuch as $B_3$ will be excluded by rule (4).  This
guarantees that $B_2$ will be stochastically chosen causing it to become conscious, thereby insuring that the
experimenter can measure $t_2$.  To the observer, $t_2$ begins another cycle of observation in which he is guaranteed that
he will become conscious of $B_3$, thereby insuring that he can measure $t_3$, etc.  

It is apparent that the theoretical restraints imposed by rule (4) are mimicked by an experimental procedure that can be
used to confirm the predictions of standard quantum theory.  The experimenter starts the experiment over each time he
becomes conscious of a new capture, treating each capture like the beginning of a new `renormalized' decay.  Collecting
ensembles of data of this kind, he can find the decay curve of each component in eq.\ 2, as well as the distribution of
counts at some final time $t$.  So a single experiment confirms the decay curves of each of the cycles of measurement
generated by rule (4); and at the same time, it confirms the standard quantum mechanical count distribution at time $t$ in
eq.\ 3.  The statistics predicted by standard quantum theory must therefore be the same as the statistics observed by a
conscious observer who functions under rules (1-4).

\section*{The General Case}

Let the initial state $S_0$ of a more general system evolve in time to give
\begin{equation}
\Phi(t \ge t_0) = S_0 + S_1 + S_2 + . . + S_m
\end{equation}
where the Hamiltonian connects adjacent components such as $S_0$ to $S_1$, and $S_1$ to $S_2$, but provides no direct link
between non-adjacent components.  In these circumstances, the initial current from $S_0$ will go exclusively into $S_1$. 
It is only after $S_1$ has gained some amplitude that current can begin to flow into $S_2$, etc.   

When a detector and a conscious observer are entangled with these components, we get
\begin{displaymath}
\Phi(t_{sc1} > t \ge t_0) = S_0D_0\underline{B}_0 + S_1D_1B_1 + S_2D_2B_2 + . . + S_mD_mB_m
\end{displaymath}
When rule (4) is added, components higher than $S_1$ are no longer in the picture prior to a stochastic hit on $S_1$ at
time $t_{sc1}$ (i.e., time of the first particle capture). so this becomes
\begin{equation}
\Phi(t_{sc1} > t \ge t_0) = S_0D_0\underline{B}_0 + S_1D_1B_1 
\end{equation}
When a stochastic hit occurs on the ready state $B_1$, there will be a state reduction giving 
\begin{equation}
\Phi(t_{sc2} > t \ge t_{sc1}) = S_1D_1\underline{B}_1 + S_2D_2B_2 
\end{equation}
and when there is another hit on $S_2$ at time $t_{sc2}$, there will be another reduction 
\begin{equation}
\Phi(t_{sc3} > t \ge t_{sc2}) = S_2D_2\underline{B}_2 + S_3D_3B_3 
\end{equation}
This process will continue, resulting in as many separate equations as there are components.

\section*{Empirical Verification of Eq.\ 6}

We empirically verify eq.\ 6 in the same way that we verified eq.\ 3.  The observer begins at time $t_0$.  At time
$t_{sc1}$ when he is first conscious of $B_1$, he zeros his clock so that it will record the time $t_1$ between $t_0$ and
$t_{sc1}$
\begin{displaymath}
t_1 = t_{sc1} - t_0
\end{displaymath}
At the time of $t_{sc2}$ when he is first conscious of $\underline{B}_2$, he again zeros his clock so that it will record
the time
$t_2$ between $t_{sc1}$ and $t_{sc2}$
\begin{displaymath}
t_2 = t_{sc2} - t_{sc1}
\end{displaymath}
This process is continued until a time $t_m$ has been reached such that 
\begin{displaymath}
\Sigma_mt_m \le t <\Sigma_mt_m + t_{m+1}\hspace{1.3cm}\mbox{where} \hspace{.3cm}
t_m = t_{sc(m)} - t_{sc(m - 1)}
\end{displaymath}
The system will then be in a state m at time $t$.  Repeat this procedure many times, establishing the distribution of
states that can be found at time $t$.

The times $t_1$, $t_2$ represent the duration of each of the equations in eqs.\ 7 and 8, and $t_m$ is the duration of the
$m^{th}$ equation generated by the rules.  It is possible to experimentally determine the decay curve for each time
interval $t_m$.  Using this information, one can determine the distribution of counts at a final time $t$, and verify that
result experimentally.  Presumably, this will confirm the final count distribution that is predicted by standard (i.e.,
Born rule) quantum theory.  Therefore, the statistical predictions of standard quantum theory will be consistent with the
statistical predictions that follow from the decay curves of each $t_m$ taken separately and sequentially, as mandated by
rule (4).

\end{document}